\documentstyle[12pt,aps,fleqn]{revtex}
\begin{document}
\bibliographystyle{prsty}
\newcommand{\C}{\cite}
\newcommand{\beq}{\begin{equation}}
\newcommand{\eeq}{\end{equation}}
\newcommand{\bea}{\begin{eqnarray}}
\newcommand{\eea}{\end{eqnarray}}
\newcommand{\bit}{\begin{itemize}}
\newcommand{\eit}{\end{itemize}}
\newcommand{\ms}{m_{\rm s}}
\newcommand{\MeV}{{\rm MeV}}
\newcommand{\fm}{{\rm fm}}
\newcommand{\alf}{{\bar a}}
\newcommand{\bet}{{\bar b}}
\newcommand{\lb}{\hfil\break }
\newcommand{\qeq}[1]{eq.\ (\ref{#1})}
\newcommand{\dsl}{ \rlap{/}{\partial} }
\newcommand{\pst}{ \rlap{/}{p}  }
\newcommand{\half}{\frac{1}{2}}
\newcommand{\quref}[1]{\cite{bolo:#1}}
\newcommand{\qref}[1]{Ref.\ \cite{bolo:#1}}
\newcommand{\queq}[1]{(\ref{#1})}
\newcommand{\qtab}[1]{Tab. \ref{#1}}
\newcommand{\wu}{\sqrt{3}}
\newcommand{\nn}{\nonumber \\ }
\newcommand{\Y}{\ {\cal Y}}
\newcommand{\Sp}{{\rm Sp\ } }
\newcommand{\Spto}{{\rm Sp_{(to)}\ } }
\newcommand{\Tr}{{\rm Tr\ } }
\newcommand{\tr}{{\rm tr\ } }
\newcommand{\sign}{{\rm sign} }
\newcommand{\linie}{\ \vrule height 14pt depth 7pt \ }
\newcommand{\intT}{\int_{-T/2}^{T/2} }
\newcommand{\Nc}{N_{\rm c}}
\newcommand{\Ne}{$N_{\rm c}$}
\newcommand{\gs}{$g_{\rm A}^{(0)}$}
\newcommand{\gt}{$g_{\rm A}^{(3)}$}
\newcommand{\go}{$g_{\rm A}^{(8)}$}
\newcommand{\Gs}{g_{\rm A}^{(0)}}
\newcommand{\Gt}{g_{\rm A}^{(3)}}
\newcommand{\Go}{g_{\rm A}^{(8)}}
\newcommand{\vx}{{\vec x}}
\newcommand{\vy}{{\vec y}}
\setlength{\mathindent}{0cm}
\pagestyle{empty}
\phantom{a}
\vskip2cm
\noindent{\large HYPERONS IN EFFECTIVE CHIRAL QUARK MODELS}
\hfill  \\

\noindent{A.
Blotz$^a$\footnote{ 
\indent Talk presented at the Conference on Hypernuclear and Strange Particle Physics
(HYP94), Vancouver, B.C., Canada, 4.-8- Juli 1994   }
M. Praszalowicz$^b$  and
K. Goeke$^a$ \hfill  \\

\noindent $^a$Institute for Theoretical Physics II, Ruhr-University
Bochum, D-44780 Bochum, \\  Federal Republic of Germany
\hfill\\

\noindent $^b$Jagellonian University, Reymonta 4, 30-059, Krakow, Poland
\hfill\\
\medskip

\indent
Baryonic correlation functions are calculated
within an effective chiral quark model
motivated by the
instanton liquid model of QCD.
Using a flavour
SU(3) symmetry  for the local four-quark interaction
and taking rotational zero modes for the quantization into account
low-energy
hadronic  observables  are in surprising agreement with experimental
data. \hfill\\
\hfill\\
\hfill\\
\noindent{\bf 1. INTRODUCTION}\\
\hfill\\
\indent Today Quantum Chromodynamics
(QCD) is believed to be the theory of the
strong interaction, though it is up to now not possible to calculate
mesonic and baryonic properties directly.
But it is believed that the classical solutions of the Yang-Mills field
equations
describing semiclassical tunneling events play an important role for the
QCD vacuum structure\cite{bolo:bpst,bolo:hooft76}.
As a consequence of the zero mode solution of the Dirac
equation in the presence of the instanton fields
effective
instanton-induced interactions between the quarks emerge\quref{hooft76}.
Within the Instanton Liquid Model \quref{dipepo} it has been shown that
the instanton dependent quark coupling constant provides - via a
momentum dependent effective quark mass - a natural ultraviolet cutoff.
However approximating the instanton dependent coupling  by
a constant one immediately obtains
quark theories similiar to the Nambu--Jona-Lasinio (NJL)
model\quref{njl1}
\beq  {\cal L}_{NJL}  ={\bar q}\left(-i\gamma_\mu\partial_\mu
      + m\right) q  + {G\over 2}  \sum_{a=0}^8
       \left[ ({\bar q}\lambda_a q)^2
           +  ({\bar q}\lambda_ai\gamma_5 q)^2   \right]
\eeq
or in bosonized form to the chiral quark model of Diakonov and Petrov
\quref{dipepo}.
Though confinement is lost in these approximations, the maybe
most important feature of QCD at low energies, i.e.
spontaneous breaking of chiral symmetry,  is maintained.
This has the desired property that the lowest bound states out of two
quarks, the pseudoscalar pions and kaons, appear as Goldstone bosons
of the broken chiral symmetry. As a result important low-energy theorems
such as PCAC and the Goldberger-Treiman relation are fulfilled.
Within these effective quark theories and taking the large $N_c$-limit
baryons
emerge as solitons which are bound states of valence quarks coupled to
the polarized Dirac sea of quarks and antiquarks. However, due to the
underlying spherically symmetric hedgehog Ansatz for the chiral fields,
these solitons are only mean-field
solutions of time-independent field configurations carrying
unit winding number.
\hfill\\
\hfill\\
\hfill\\
\noindent{\bf 2. HYPERON MASS SPLITTINGS}\\
\hfill\\
The starting point for the calculation of hadronic properties is
the point-to-point correlation function
${\cal C}_h(T)=
\langle Q_h(\vx_0,T/2) Q_h^\dagger(\vy_0,-T/2)\rangle$, which in
the case of baryons is defined as the expectation value of the Ioffe
currents
\beq
Q_B(\vx_0,t)={1\over N_c!} \epsilon^{\alpha_1\dots\alpha_{N_c}}
\Gamma^{f_1\dots f_{N_c}}q(x)_{f_1}\dots q(x)_{f_{N_c}}  \eeq
where  $\Gamma^{f_1\dots f_{N_c}}$ is a symmetric matrix in flavour and
spin space and $\epsilon^{\alpha_1\dots\alpha_{N_c}}$ is a total
antisymmetric tensor with respect to colour.
Then for large Euclidean times T it reduces to \quref{dipepo}
\beq   {\cal C}_B(T) =
        \stackrel{\rm T\to \infty }{\simeq}
        \Gamma^{f_1\dots f_{N_c}}  \Gamma^{g_1\dots g_{N_c}*}
          \Pi_{i=1}^{N_c} \left[ \phi_{n,f_i}(\vx_0)
                              \phi_{n,f_i}^\dagger(\vy_0) \right]
         e^{- T E[B=1] }
\eeq
where  $E[B=1]$ is the classical mean-field energy.
In order to describe states with the quantum numbers of the physical
baryons a semiclassical quantization scheme is adopted \quref{anw}.
In this scheme the rotational zero-modes of the one-loop effective
action are quantized
assuming an approximate symmetry of flavour SU(3).
Hyperon masses follow then again from the behaviour of the
corresponding correlation functions in the large Euclidean time limit.
As a result octet [8] and decuplet [10]
representations for the hyperons appear as lowest possible states.
Incorporation of symmetry breaking effects via strange and
non-strange current quark masses yields then a nice explanation
of physical spectra.
To be precise hyperon splittings are determined
within $\pm 10\MeV$ accuracy\quref{ab4}, whereas isospin splittings
(especially n-p)
are reproduced even within the experimental error bars\quref{ab5}. One
should
stress that the parameters of the model are strictly fixed by requiring
proper pion-decay and pion and kaon masses\quref{ab4}. The
constituent quark mass M is fixed in the baryon sector
to be $M\simeq 420\MeV$ reproducing the hyperon spectra for
$m_s\simeq 180\MeV$ (cf. Fig.1).
\hfill\\
\hfill\\
\hfill\\
\noindent{\bf 3. AXIAL CURRENTS}\\
\hfill\\
The recent EMC measurements on the nucleon structure
functions suggested that only a small fraction of the proton spin
(cf. $g_A^{(0)}$ in Tab.1) is
carried by the spins of quarks.  This is in drastic contradiction to the
naive quark model, in which the spins of $N_c$ quarks are coupled to the
known spin of the proton, and is therefore denoted as {\it spin
crisis}.
In the present model it was shown\quref{ab6} that the experimental
number\quref{elka2} for $g_A^{(0)}$ can be explained with only a
moderate contribution
of the strange quarks ($\Delta s$ in Tab.1). Furthermore the $g_A^{(3)}$
and $g_A^{(8)}$ are evaluated\quref{ab9} (cf. Fig.2).
The reason that the present values  are in better agreement with
experiment compared to former Skyrme model calculations
is
the next to leading order rotational correction in
the $1/N_c$-expansion.  These subleading terms are entirely due to the
non-local
structure of the effective action and cannot be obtained in
local effective meson theories.
However qualitatively they agree with recent large-$N_c$ estimates
of Dashen and Manohar\quref{dama2}, which state that the coupling
constant ratio of different hyperons obtains only corrections at the
level of $1/N_c^2$. Numerically the nucleonic $g_A^{(3)}$, which was
evaluated as $\simeq 0.8$ in most of the chiral models,
is now close to the experimental value because of the
subleading $1/N_c$ corrections (cf. Tab.1).
\hfill\\
\begin{figure}[t]
\unitlength1cm
\begin{center}
\begin{minipage}{16.0cm}
\begin{picture}(16.0,7.0)
\put(0.0,0.0){\framebox(7.8,7){ } }
\put(8.2,0.0){\framebox(7.8,7){}  }
\end{picture}
\smallskip
\parbox[t]{7.5cm}{
Figure 1. The deviations of the theoretical from the experimental mass
for the [8] and [10] baryons as a function of $m_s$.
}
\hfill
\parbox[t]{7.5cm}{
Figure 2. The axial vector coupling constants $g_A^{(0)}$, $g_A^{(3)}$
and $g_A^{(8)}$
as a function of the constituent quark mass  M. }
\end{minipage}
\end{center}
\end{figure}
\hfill\\
\hfill\\
\noindent{\bf 4. OTHER OBSERVABLES  }\\
\hfill\\
In addition we considered strangeness corrections to the pion and kaon
nucleon
$\sigma_{\pi N,KN}$-terms, which are
related to the formfactor of meson-nucleon scattering amplitudes
at low $q^2$.
Whereas $\sigma_{\pi N}$ coincides with the experimental
data\quref{galesa}
(cf. Tab.1),  $\sigma_{KN}$ is experimentally not
yet known. Furthermore
the Gottfried sum measured by NMC\quref{nmc}, which states that
\beq  S_G = {1\over 3} +{2\over 3} \int dx  \left(
         {\bar u}_{sea}^p - {\bar d}_{sea}^p \right)
         = 0.240  \ne {1\over 3}
\eeq
where ${\bar q}_{sea}^p$ is the quark distribution function
in the parton model within the proton (p),
suggests
a flavour
asymmetric polarization of the up and down Dirac sea\quref{wa2}.
This effect is qualitatively as well as quantitatively
reproduced\quref{blotz} (cf.
Tab.1) within the present model.
\begin{table}[h]
\noindent{Table 1}\\
The  axial vector coupling constants $g_A^{(0)},g_A^{(3)},g_A^{(8)}$,
its strangeness contribution $\Delta s$,
the pion and kaon-nucleon $\sigma$-terms,
the Gottfried sum
$S_G$ and the strange content $y=2{\bar
s}s/({\bar u}u+{\bar d}d)$ of the nucleon for the NJL model compared
with 'experimental' values.
\label{taball}
\begin{center}
\begin{tabular}{ccccccccc}
 &   $g_{\rm A}^{(0)}$  &  $g_{\rm A}^{(3)}$  &   $g_{\rm A}^{(8)}$
 & $\Delta s$ &
    $\sigma_{\pi N}[\MeV]$  &  $\sigma_{KN}[\MeV]$
  &  $S_G$  &  y        \\
\hline
 NJL & 0.37         & 1.38 &  0.31        & -0.05  & 45.2   & 499  &
         0.234      &  0.457 \\
 exp & 0.31$\pm$0.07 & 1.26 & 0.35$\pm$0.04 & -0.10 & 45$\pm$8 & ? &
         0.24$\pm$0.016  & ? \\
\end{tabular}
\end{center}
\end{table}
\noindent{\bf 5. SUMMARY  }\\
\hfill\\
Starting from an instanton liquid motivated
effective quark interaction with SU(3) flavour symmetry it has been
shown that low-energy baryonic observables, such as hyperon and
isospin
mass splittings, $\sigma$-terms  and axial vector coupling constants,
are in surprising agreement with experiment.
Simultaneously the
parameters of the theory are fixed by the masses and decay constants of
the almost Goldstone bosons.
This suggests that
chiral symmetry and the
spontaneous breaking of this symmetry is a key ingredient of an
effective theory for the hadrons.


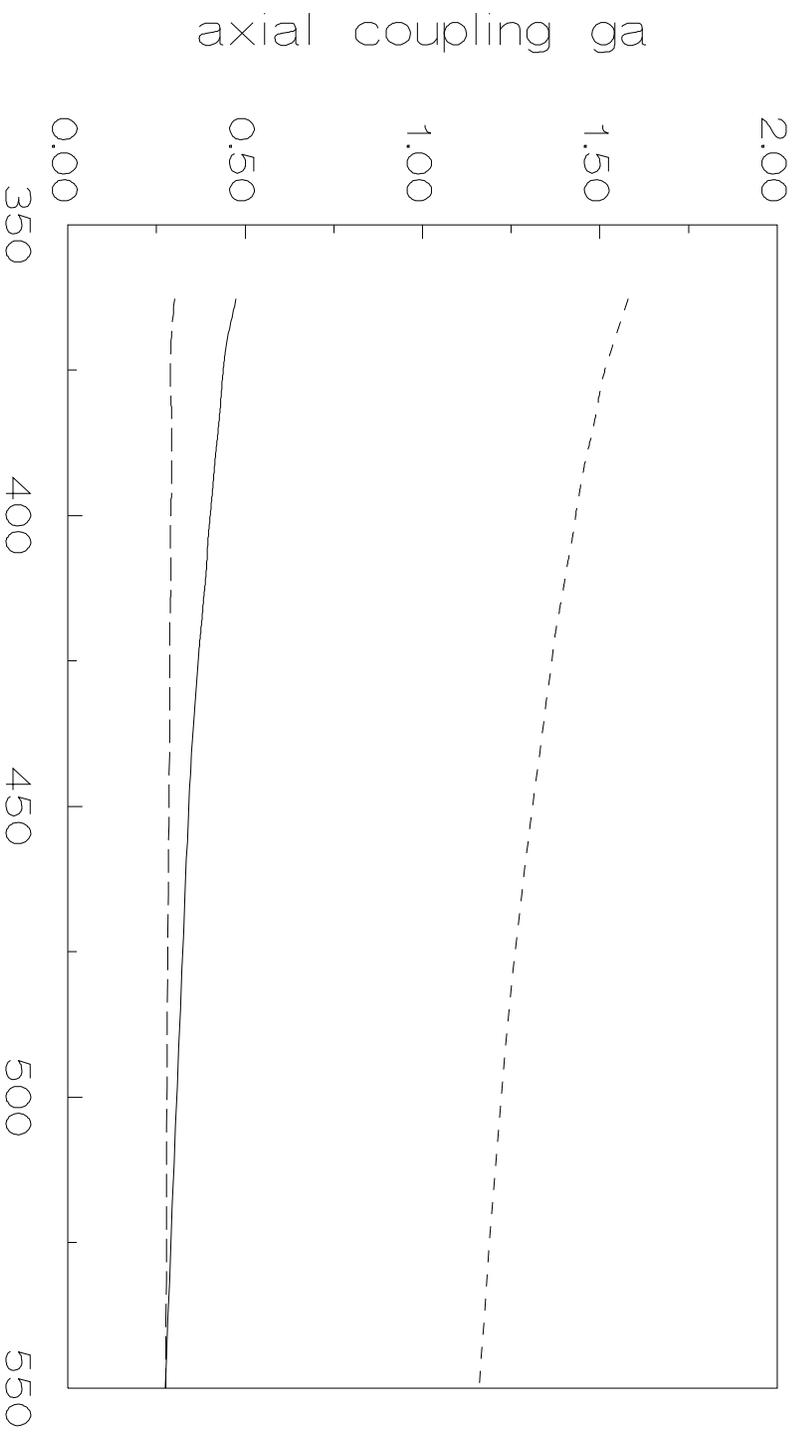
\end{document}